\documentclass[12pt]{article}
\usepackage{graphics,epsf,epsfig}
\usepackage{amssymb}

\begin{document}

\title{Motion of a transverse/parallel grain boundary in a block
copolymer under oscillatory shear flow}
\author{Fran\c{c}ois Drolet and Jorge Vi\~nals$^*$ \\
School of Computational Science and Information Technology \\
Florida State University, Tallahassee, Florida 32306-4120}
 
\maketitle

\begin{abstract}

A mesoscopic model of a diblock copolymer is used to study the motion
of a grain boundary separating two regions of perfectly ordered
lamellar structures under an oscillatory but uniform shear flow. The case 
considered is a grain boundary separating lamellae along the so called 
parallel orientation (with wavevector parallel to the velocity gradient 
direction) and along the
transverse orientation (wavevector parallel to the shear
direction). In the model considered 
lamellae in the parallel orientation are marginal with
respect to the shear, whereas transverse lamellae are
uniformly compressed instead. A multiple scale expansion valid in
the weak segregation regime and for low shear frequencies leads to an
envelope equation for the grain boundary. This equation shows that
the grain boundary moves by the action of the shear, with a
velocity that has two contributions. The first contribution, which
arises from the change in orientation of the grain boundary as it is being
sheared, describes variations in monomer density  in the 
boundary region. This contribution is periodic in time with
twice the frequency of the shear and, by itself, leads to no net motion of the
boundary. The second contribution arises from a free energy increase
within the bulk transverse lamellae under shear. It is also periodic with
twice the frequency of the shear, but it does not average to zero,
leading to a net reduction in the extent of the region occupied by the 
transverse lamellae.  We find in the limit studied that
the velocity of the grain boundary can be expressed as
the product of a mobility coefficient times a driving force, the
latter given by the excess energy stored in the transverse phase being sheared.

\end{abstract}

\section{Introduction}
\label{sec:introduction}

Self-assembly of block copolymers is one possible route to the
development of nanostructured materials, either directly or as
templates. The major challenge that needs to be overcome for widespread 
application of these materials is
the development of long ranged order in the polymer
over scales much larger than the
wavelength of the mesophase. The purpose of this paper is to
investigate the motion of a grain boundary, an extended topological
defect separating two lamellar phases that exhibit long ranged
order but along different directions. Our study is an extension of ref.
\cite{re:boyer01} in that we explicitly consider here that the block 
copolymer is
under an externally imposed, oscillatory and uniform shear flow. The
study is motivated by the widespread use of both steady and oscillatory
shear flows to induce long ranged order in lamellar phases.

The system under consideration is a symmetric diblock copolymer slightly
below its order-disorder transition temperature $T_{ODT}$ (weak segregation
regime). The equilibrium phase is a lamellar structure in
which nanometer sized layers rich in A or B monomers
alternate in space. When the copolymer is quenched from a high
temperature to a temperature $T < T_{ODT}$, a transient polycrystalline
sample results comprised of an ensemble of locally ordered grains but of
arbitrary orientations. A large number of defects are present in the
sample in addition to grain boundaries that include dislocations and
disclinations.

Different methods of sample alignment are being investigated
experimentally, including substrate induced patterning 
\cite{re:fasolka97,re:rockford99,re:fasolka01}, 
step induced 
orientation of thin films \cite{re:segalman01}, electric fields that take 
advantage of a non uniform dielectric constant
\cite{re:amundson93,re:amundson94}, or of the existence of ions
in the copolymer \cite{re:tsori03}, and 
oscillatory shear flows in bulk samples
\cite{re:hadziiaannou79,re:koppi92,re:koppi93,re:patel95,re:gupta95,%
re:leist99}, We focus here on the latter case as there
is no agreement at present on the issue of orientation selection as a
function of the physical properties of the copolymer and the parameters
of the flow. For the purposes of the discussion, three basic
orientations of the lamellae relative to the shear flow are conventionally 
defined: parallel, in which the 
lamellar planes are parallel to the flow velocity; transverse in which
the lamellar normal is parallel to the flow, and perpendicular in which
the lamellar normal is parallel to the vorticity of the imposed flow. A
review of current experimental phenomenology can be found in
\cite{re:hamley98,re:larson99}..

In an attempt to clarify the existing phenomenology about orientation
selection of lamellar phases, we recently undertook a theoretical
stability analysis of such phases under oscillatory shear flows with
and without
viscosity contrast between the microphases \cite{re:drolet99,re:chen02}.
At fixed temperature, we found that there is a finite range of shear 
amplitudes within which periodic lamellar structures along a given
direction exist. For amplitudes larger than a certain critical amplitude, the
lamellar phase \lq\lq melts" into a disordered phase, possibly reforming
with a different orientation relative to the shear, the new 
orientation being within the band of
allowed solutions. Lamellar configurations within the band of allowed
solutions can in turn become unstable against long wavelength
perturbations. The regions of occurrence of this secondary instability 
were given in refs.
\cite{re:drolet99,re:chen02} as a function of the orientation of the lamellae 
and the shear rate. Generally speaking, it was found that the region of
stability of perpendicular lamellae is larger than that of parallel
lamellae, and both considerably larger than the region of stability of
the transverse orientation. Our results also showed that the critical mode of
instability is typically along the perpendicular
orientation, so that the decay of an unstable region of parallel or
transverse lamellae would lead, at least initially, to lamellae
predominantly oriented along the perpendicular direction. These results
were interpreted through a geometric description of the lamellar distortion, 
suggesting 
that the emerging mode of instability is the one that causes the
largest decrease in lamellar wavelength under shear. Finally, the
results were shown to be fairly insensitive to viscosity
contrast between the microphases.

While the results just summarized narrow the range over which particular
orientations can be in principle observed experimentally, they do not 
provide an orientation
selection mechanism among competing, linearly stable stationary states. 
We therefore turn our attention to dynamical aspects of the competition
between 
coexisting orientations in a macroscopic sample, and to
orientation selection mechanisms of dynamical nature. We focus in this
paper on the motion of a front or boundary that separates two ordered
regions of parallel and transverse orientations. Parallel lamellae in the
mesoscopic description employed are marginal with respect to the shear,
and therefore unaffected by the flow. Transverse lamellae, on the other
hand, are compressed by the shear, a fact that is shown to induce
boundary motion toward the region of transverse orientation. 
Therefore parallel lamellae are expected to become prevalent over
transverse lamellae, even in
those ranges of parameters of the polymer and of the flow in which
transverse lamellae are linearly stable.

The role that topological defect motion plays on structure coarsening
under shear flows has already been investigated numerically in 
\cite{re:zvelindovsky98,re:ren01}. Either a density functional
description (\cite{re:zvelindovsky98}), or a cell dynamical system model
(\cite{re:ren01}) have been used to study domain coarsening of an
initially macroscopically disordered configuration. For {\em steady} shears,
Zvelindovsky et al. \cite{re:zvelindovsky98} show that the shear is
very effective in speeding up the formation of lamellar domains, and
argue that the perpendicular orientation is most stable. Ren et al. 
\cite{re:ren01} more specifically focused on topological defect motion
and annihilation, and the amplitudes of the oscillatory shear
necessary to eliminate them to form a perfectly ordered lamellar structure.

\section{Mesoscopic model equation of a lamellar phase under shear}
\label{sec:model}

At a mesoscopic level, and for time scales that are long compared with the
relaxation time of the polymer chain, a block copolymer melt is 
described by an order
parameter $\psi({\bf r})$ which represents the local density
difference between the two monomers constituting the copolymer. The
corresponding free energy was derived by Leibler 
in the weak segregation limit (close to $T_{ODT}$) \cite{re:leibler80}, and 
later extended by Ohta and Kawasaki to the strong segregation
regime \cite{re:ohta86}. If the temporal evolution of $\psi$ occurs through 
advection by a flow field as well as through local dissipation driven by free
energy reduction, $\psi$ obeys a time dependent Ginzburg-Landau
equation that in the symmetric case of equal volume fraction of the
two monomers is given by \cite{re:oono88b},
\begin{equation}
\frac{\partial \psi}{\partial t} + {\mathbf v} \cdot \nabla \psi = 
\nabla^{2} \left( - \psi + \psi^{3} - \nabla^{2} \psi \right) - B \psi.
\label{eq:psieq}
\end{equation}
All quantities have been made dimensionless, including the advection
velocity ${\bf v}$, and the long range polymer interaction coefficient
$B$. The order-disorder transition between a disordered phase ($\psi = 0
$) and a lamellar phase ($\psi \neq 0$) takes place at $B_{0}= 1/4$.
For $B \gtrsim B_{0}$, $\psi$ is a periodic function of wavenumber 
$q_{0} = 1/\sqrt{2}$.

\begin{figure}
\centerline{\epsfig{figure=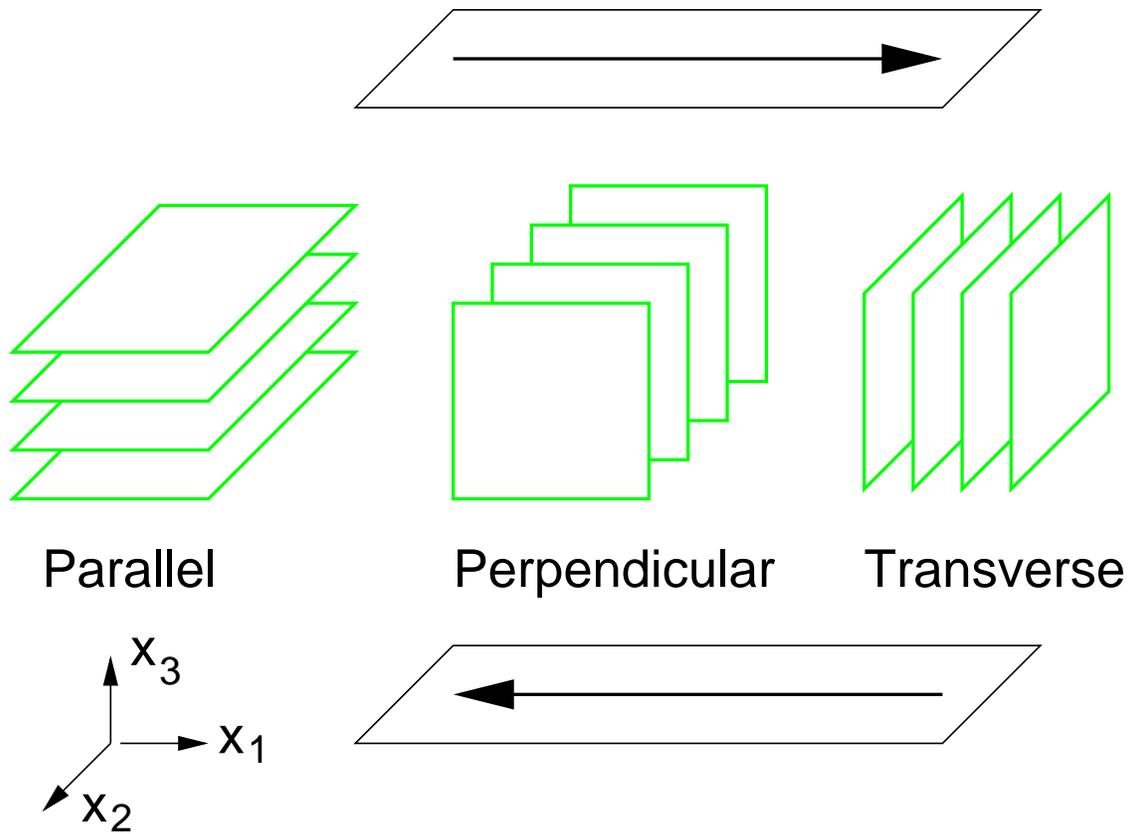,width=6in}}
\caption{Schematic representation of the geometry considered including
the shear direction, and the three different lamellar orientations discussed
in the text.}
\label{fig:alignment}
\end{figure}

The physical system under consideration here is a layer of block copolymer,
unbounded in the $x$ and $y$ directions, and being uniformly
sheared along the $z$ direction (Fig. \ref{fig:alignment}). The
layer is confined between the stationary $z=0$ plane, and the
plane $z=d$ which is uniformly displaced parallel to itself with a
velocity ${\mathbf v}_{\mathrm{plane}} = \gamma \, d \, \omega \,
\cos(\omega t) \hat{\bf x}$, where $\hat{\bf x}$ is the unit
vector in the $x$ direction.

We first briefly summarize the results of refs. 
\cite{re:drolet99,re:chen02} concerning
stationary lamellar solutions in shear
flow. In the weak segregation limit $\epsilon = (B - B_{0})/2B_{0}
\ll 1$, the solution for the monomer composition can be obtained 
perturbatively in $\epsilon$,
\begin{equation}
\psi ({\mathbf r}) = 2 A(t) \cos ({\mathbf q} \cdot {\mathbf r}) +
A_{1}(t) \cos ( 3 {\mathbf q} \cdot {\mathbf r} ) + \ldots
\label{eq:sol2}
\end{equation}
where $A(t) \sim {\cal O}(\epsilon^{1/2})$, and $A_{1}(t)$ and higher
order mode amplitudes are of higher order in $\epsilon$. We have
defined ${\mathbf r} = x_{1} \hat{\bf x} + x_{2} \hat{\bf y} +
x_{3} (\gamma \sin (\omega t) \hat{\bf x} + \hat{\bf z})$, a vector that is
expressed in a non orthogonal basis 
set which follows the
imposed shear, and ${\mathbf q}=(q_{1}, q_{2}, q_{3})$, the
wavevector in the corresponding reciprocal space basis set $\{
{\mathbf g}_{1} = \hat{\bf x} - \gamma \sin (\omega t) \hat{\bf
z}, {\mathbf g}_{2} = \hat{\bf y}, {\mathbf g}_{3} = \hat{\bf
z} \}$. Note that in this new coordinate system, the wavevector of a 
perfectly ordered configuration is stationary.
Three orientations relative to the
shear can be defined as follows: $q_{3} \neq 0, q_{1}=q_{2} = 0$ is a
purely parallel orientation, $q_{2} \neq 0, q_{1}=q_{3}=0$ is a
perpendicular orientation, and $q_{1} \neq 0, q_{2}=q_{3}=0$ is a
transverse orientation.

For constant viscosity and if we neglect flow induced by the lamellae 
themselves, the velocity field in the layer is independent
of monomer composition, and is given by,
\begin{equation}
{\mathbf v} = \gamma \omega \cos (\omega t) \; z \; \hat{\bf x}.
\label{eq:basev}
\end{equation}
To lowest order in $\epsilon$, the amplitude $A(t)$ satisfies the equation 
\cite{re:drolet99},
\begin{equation}
\frac{dA}{dt} = \sigma[q^2(t)]A - 3 q^2(t)A^3,
\label{eq:ampl}
\end{equation}
with $q^2(t)=q_1^2 + [\gamma \sin (\omega t) q_1 -q_3]^2 + q_2^2$ and
$\sigma(q^2)=q^2-q^4-B$.  This equation can be integrated to give the
marginal stability boundaries, and the function $A(t)$ itself
\cite{re:drolet99}. From this analysis, a critical strain amplitude
$\gamma_{c}$ was identified, function of the orientation ${\bf q}$ but
independent of the frequency $\omega$,
such that for $\gamma < \gamma_{c}$ the
uniform lamellar structure oscillates with the imposed shear, but for
$\gamma > \gamma_{c}$ $A(t)$ decays to zero; i.e., the lamellar
structure melts, according to the terminology used by experimentalists.

The stability of this base lamellar pattern was then addressed by
Floquet analysis. Regions of stability were obtained for lamellar solutions of 
arbitrary orientation, that were generally largest for orientations
near the perpendicular direction, and smallest in the vicinity of the
transverse direction. As discussed in the introduction,
this stability analysis provides some guidance on the issue of orientation
selection, but we wish to extend here the analysis of existence and stability
to possible selection by dynamical mechanisms. The specific
case considered in this
paper is the motion of a grain boundary separating regions
of uniform parallel and transverse orientations under oscillatory shear.

We use in what follows a different form of the equation
governing the evolution of the monomer composition $\psi$, known as
the Brazovski equation (or Swift-Hohenberg equation in the fluids 
literature) \cite{re:swift77,re:cates89,re:fredrickson94}.
Both this equation and eq (\ref{eq:psieq}) lead to the same amplitude or
envelope equations near onset \cite{re:shiwa97,re:boyer02}, and hence
lead to identical results in the limit addressed in this paper. The
Swift-Hohenberg equation for a scalar order parameter is,
\begin{equation}
\frac{\partial \psi }{\partial t} + {\bf{v}} \cdot {\bf{\nabla}} \psi= 
\epsilon \psi - (\nabla^2 + q_0^2)^2 \psi - \psi^3,
\label{eq:she}
\end{equation}
where all quantities are dimensionless. In this units $q_{0} = 1$, although 
we will retain the symbol $q_{0}$ in the equations that follow for clarity
of presentation.

As was the case in the analyses presented in refs.
\cite{re:drolet99,re:chen02}, we introduce a new frame of
reference in which the velocity vanishes. In the case of an imposed oscillatory
shear of  amplitude  $\gamma$ and angular frequency $\omega$, we define a
set of non-orthogonal coordinates $x'=x- a(t) z$ and  $z'=z$ where 
$a(t) = \gamma \sin (\omega t)$. We assume that the system is uniform
in the third direction, and therefore simply focus on a two dimensional case.
Equation (\ref{eq:she}) transforms to,
\begin{equation}
\frac{\partial \psi }{\partial t} = \epsilon \psi -
 (\nabla'^2 + q_0^2)^2 \psi - \psi^3,      
\label{eq:shep}
\end{equation}
where
\begin{displaymath}
\nabla'^2 = (1 +  a(t)^2) \frac{\partial^2}{\partial x'^2} -2 a(t)
\frac{\partial^2}{\partial x' \partial z'} + \frac{\partial^2}{\partial z'^2}.
\end{displaymath} 

A solution of the linearization of eq (\ref{eq:shep}) can be found by 
assuming
$$
\psi ({\mathbf r}^{\prime}) = A \cos \left( {\mathbf q}^{\prime} \cdot
{\mathbf r}^{\prime} \right).
$$
Given that $\left( \nabla^{~\prime 2} + q_{0}^{2} \right) \psi = \left(
-q(t)^{2} + q_{0}^{2} \right) \psi$ with $q(t)^{2} = q_{x'}^{2} + 
\left( a(t) q_{x'} - q_{z'} \right)^{2}$, we find
\begin{equation}
\frac{dA}{dt} = \epsilon A - \left( -q(t)^{2} + q_{0}^{2} \right)^{2} A =
\sigma(t) A.
\label{eq:linearA}
\end{equation}
The disordered solution $A = 0$ becomes unstable when
$$
\int_{0}^{2\pi/\omega} \sigma(t')dt' > 0.
$$
In analogy with the case analyzed in ref. \cite{re:drolet99}, we find several
instability modes and associated thresholds,
$$
\begin{array}{lll}
\bullet ~ q_{x} = 0, q_{z}=q_{0} & \epsilon = 0 & {\rm parallel~mode} \\
\bullet ~ q_{z} = 0, q_{x} = \sqrt{\frac{4\gamma^{2} + 8}{3\gamma^{4} + 
8 \gamma^{2} + 
8}} \; q_{0} & \epsilon = \frac{\gamma^{4} q_{0}^{4}}{3\gamma^{4} + 8 
\gamma^{2} + 8} & {\rm transverse~mode} \\
\bullet ~ q_{x} = \sqrt{\frac{2}{15\gamma^{2} + 16}} \; q_{0}, 
\quad q_{z} = 
\sqrt{\frac{3 \gamma^{2} +8 }{15\gamma^{2} + 16}} \; q_{0} &
\epsilon = \frac{8 \gamma^{2} q_{0}^{4}}{15 \gamma^{2} + 16} & 
{\rm mixed~mode}
\end{array}
$$

\section{Envelope equation for a grain boundary under weak shear}
\label{sec:envelope}

\begin{figure}
\centerline{\epsfig{figure=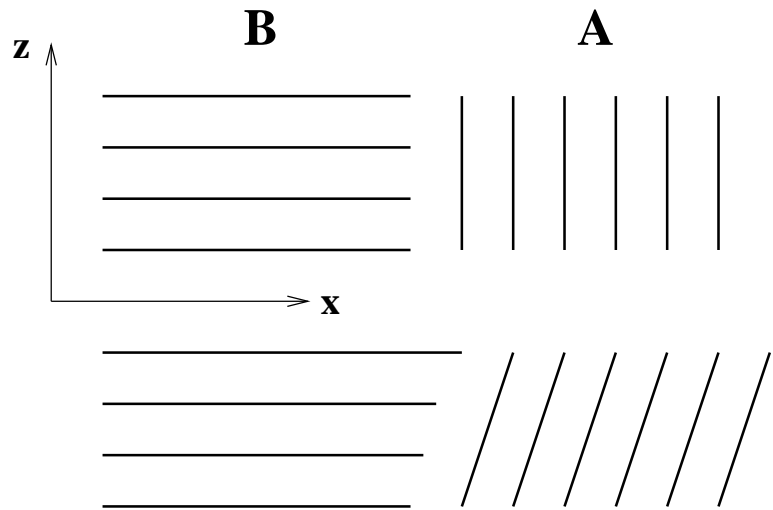,width=4in}}
\caption{Schematic representation of a planar grain boundary that separates
regions of parallel and transverse lamellae being uniformly sheared.}
\label{fi:gb_schem}
\end{figure}

We focus on a special configuration comprising two perfectly ordered 
lamellar domains,
initially oriented perpendicular to each other that meet at a 
grain boundary. We assume that both domains are initially of same
wavenumber $q_0$ at least far from the boundary, so that a planar
grain boundary would be stationary in the absence of shear.
We will neglect in this study any back flow induced by the lamellae
themselves (through osmotic stresses), so that 
the velocity field ${\bf{v}}$ in eq (\ref{eq:she}) equals 
the imposed shear flow. A schematic representation of the 
configuration under study is shown in Fig. \ref{fi:gb_schem}. 
We denote by B the lamellae that lie parallel to the flow field, and note
that the order parameter $\psi$ in this region is unaffected by the flow.
Transverse lamellae are denoted by A. If the A lamellae were
to adiabatically follow the imposed flow, both orientation and
wavelength would be a periodic function of time as 
illustrated schematically on Fig. \ref{fi:gb_schem}.
Because a local change in wavenumber away from $q_{0}$ always
leads to a free energy increase in region A, while the free energy in region B 
remains unchanged, we anticipate grain boundary motion from region B to 
region A, thereby increasing the area occupied by parallel lamellae. 

We derive a set of amplitude equations from eq (\ref{eq:shep}) by
using a multiple scale approach. For $\epsilon \ll 1$, it is possible to 
extract the slow evolution of both the lamellae and the grain boundary 
(on a slow time scale $\epsilon t$) by expanding the order parameter 
$\psi$ in both regions around a periodic function, with amplitudes that
are slowly varying in the grain boundary region (of very large extent
in this limit). Our derivation follows closely that of Tesauro and Cross
for the case of no flow \cite{re:tesauro87}. 

The analysis is restricted to shears of small amplitude
and low frequency. Specifically, since $-q(t)^{2} + q_{0}^{2} =
-a^{2} q_{x'}^{2} + 2a q_{x'}q_{z'}$, the shear contributes to eq
(\ref{eq:linearA}) with two terms, one of order ${\cal O}(a^{2})$ and 
the other of order ${\cal O}(a^{4})$. Consistency with the expansion
in the weak segregation limit requires that $(-q(t)^{2}+q_{0}^{2}) 
\sim {\cal O}(\epsilon)$, a requirement that dictates the magnitude of the
shear amplitude. For the case considered below involving a grain
boundary between parallel and transverse orientations, the cross
derivative term $-2a q_{x'}q_{z'}$ acting on the reference state vanishes,
and therefore we will have $(-q(t)^{2}+q_{0}^{2})^{2} \sim
{\cal O}(a^{4})$ or $a \sim {\cal O}(\epsilon^{1/4})$.

We start by assuming that the slowly varying amplitude of
mode $e^{i q_0 x'}$ in region A has as characteristic length 
scales $X=\epsilon^{1/2} x'$ and $Z=\epsilon^{1/4} z'$, while the mode
$e^{i q_0 z'}$ in domain B has characteristic scales 
$\bar{X}=\epsilon^{1/4} x'$ and $\bar{Z}=\epsilon^{1/2} z'$. 
We further assume a weak shear so that $a \sim \epsilon^{1/4}$.
This scaling is appropriate for an initial configuration in which the
transverse lamellae are parallel to the grain boundary (and therefore
in the limit of small shear amplitude, the transverse lamellae will remain
almost perpendicular to the parallel lamellae for all times). If the 
angle between the transverse lamellae and the grain boundary
is finite, then the scaling $a \sim \epsilon^{1/2}$ needs to be 
introduced instead.
The operator  $(\nabla'^2 + q_0^2)^2$ can now be expanded in powers of
$\epsilon$ as
\begin{eqnarray}
\nonumber
 (\nabla'^2 + q_0^2)^2 &=& L_0^2 + \epsilon^{1/4} (2 L_0 L_1) + \epsilon^{1/2}
(L_1^2 + 2 L_0 L_2) \\ & & + \epsilon^{3/4} (2 L_0 L_3 + 2 L_1 L_2) + 
  \epsilon (2 L_0  L_4 + 2 L_1 L_3 + L_2^2), 
 \label{eq:oper}
\end{eqnarray}
where we have defined,
\begin{itemize}
  \item[] $L_0= \partial^2_{x'} + \partial_{z'}^{2} + q_0^2$
  \item[] $L_1= 2 (\partial_{x'} \partial_{\bar{X}} + \partial_{z'} \partial_Z
             - a \partial_{x'} \partial_{z'})$
  \item[] $L_2= \partial^2_{\bar{X}} + \partial^2_Z  + 2 (\partial_{x'}
\partial_X +  \partial_{z'} \partial_{\bar{Z}} - a  \partial_{z'} 
\partial_{\bar{X}}
- a \partial_{x'} \partial_Z) + a^2 \partial^2_{x'} $
  \item[] $L_3= 2 [\partial_{\bar{X}} \partial_X + \partial_Z
      \partial_{\bar{Z}} -a (\partial_{z'} \partial_X 
 +\partial_{x'} \partial_{\bar{Z}} + \partial_{\bar{X}} \partial_Z) 
    + a^2 \partial_{x'} \partial_{\bar{X}} ]   ~$ and
  \item[] $L_4 = \partial^2_X + \partial_{\bar{Z}}^2 - 2a( \partial_X
    \partial_Z +  \partial_{\bar{X}}  \partial_{\bar{Z}}) + a^2
    (2 \partial_{x'} \partial_X + \partial^2_{\bar{X}}).$
\end{itemize}

We also expand $\psi$ as
\begin{equation}
 \psi = \epsilon^{1/2} \psi_0 +  \epsilon \psi_1 +  \epsilon^{3/2} \psi_2 + ...
 \label{eq:exp}
\end{equation}
and assume that both the frequency of the imposed shear, and 
the associated variation of $\psi$ is over a slow time scale 
$T= \epsilon t$. From 
eqs (\ref{eq:shep}), (\ref{eq:oper}) and (\ref{eq:exp}) we obtain, 
at $O(\epsilon^{1/2})$ the equation
\begin{equation}
- L_0^2 \psi_0 = 0,
\label{eq:eps1}
\end{equation}
which admits the solution 
\begin{equation}
\psi_0 = \frac{1}{\sqrt{3}} [A_0 e^{i q_0 x'} + B_0  e^{i q_0 z'} +
\mbox{ c.c}],
\end{equation}
with $A_0$ and $B_0$ functions of $X,Z,\bar{X},\bar{Z}$ and
$T$. At $O(\epsilon)$, eq (\ref{eq:shep}) reduces to 
\begin{equation}
-L_0^2 \psi_1 = L_1^2 \psi_0,
\label{eq:eps2}
\end{equation}
where we have used the fact that $L_0 \psi_0 =  0$, and taken advantage that
the cross derivative term $a \partial_{x'}\partial_{z'}$ vanishes when
acting on $\psi_{0}$, so that the solution at this order is also time
independent (in the sheared frame of reference). Since $\psi_0$ 
is an eigenmode of $L_0$ with zero eigenvalue, the right-hand side of 
eq (\ref{eq:eps2})  must vanish in order for it to admit a solution.
Solvability requires that the scalar product 
$\langle \psi_{0}^{+} | L_{1}^{2} \psi_{0} \rangle =
0$, that is, the right hand side of eq (\ref{eq:eps2}) is orthogonal to the
the zero eigenfunctions of the adjoint of $L_{0}$. 
But $ \langle \psi_{0}^{+} | L_{1}^{2} \psi_{0} 
\rangle = \langle L_{1}^{+} \psi_{0}^{+} | L_{1} \psi_{0} \rangle =
\| L_{1} \psi_{0} \|^{2} = 0$, from which it follows $ L_{1} \psi_{0} =0$.
As a result of this solvability condition, $A_0$ and $B_0$ must 
be independent of $\bar{X}$ and $Z$, respectively. Equation 
(\ref{eq:eps2}) then  
reduces to eq (\ref{eq:eps1}) and $\psi_1=  [A_1 e^{i q_0 x'} 
+ B_1  e^{i q_0 z'} + \mbox{ c.c},]/\sqrt3$. 

Finally, at $O(\epsilon^{3/2})$, the multiple scale analysis yields the 
equation
\begin{equation}
L_0^2 \psi_2 = - \partial_T \psi_0 + \psi_0  - \psi_0^3 - (2L_0 L_4 + 2 L_1
L_3 + L_2^2) \psi_0 - (2 L_0 L_2 + L_1^2) \psi_1.
\label{eq:psi2}
\end{equation}
Again, the functions $\psi_0$ and $\psi_1$ are zero eigenmodes of the operator 
$L_0$, so that the projections of the terms in the right-hand side of 
eq (\ref{eq:psi2}) on these eigenfunctions must vanish. 
From this condition, we obtain the following amplitude equations  
\begin{equation}
\partial_T A_0 = \{ 1- [ 2 i q_0 (\partial_X -a \partial_Z) + \partial^2_Z 
 - q_0^2 a^2]^2 \} A_0 -|A_0|^2 A_0 -2 |B_0|^2 A_0,
\label{eq:A1}
\end{equation} 
and, 
\begin{equation}
\partial_T B_0 = \{ 1- [ 2 i q_0 (\partial_{\bar{Z}} -a \partial_{\bar{X}}) 
+ \partial^2_{\bar{X}} ]^2 \} B_0 -|B_0|^2 B_0 -2 |A_0|^2 B_0,
\label{eq:B1}
\end{equation} 
where we have used the fact that $L_0 \psi_0 = L_1 \psi_0 =0$ and have
set $ L_1^2 \psi_1 = 0$.
This set of equations (\ref{eq:A1}) and (\ref{eq:B1}) governs the evolution of 
the slowly varying envelopes
of the base lamellar pattern, including variations both in the direction
parallel and perpendicular to the grain boundary. We now restrict our
attention to the case of a planar grain boundary; hence we
do not consider any dependence of the amplitudes on the coordinate parallel 
to the grain boundary. Transverse perturbations of the grain boundary are
expected to decay back to planarity, and such decay will not be
considered here. Therefore $A_0$ and $B_0$ depend only on $X$ and $\bar{X}$ 
respectively. Equations (\ref{eq:A1}) and (\ref{eq:B1}) simplify to    
\begin{equation}
\partial_T A_0 = [ 1- ( 2 i q_0 \partial_X - q_0^2 a^2)^2 ] A_0 -|A_0|^2 A_0 
-2 |B_0|^2 A_0,
\label{eq:A}
\end{equation}      
and,
\begin{equation}
\partial_T B_0 = [ 1- (\partial^2_{\bar{X}} - 2 i a q_0 
\partial_{\bar{X}})^2 ] B_0 -|B_0|^2 B_0 -2 |A_0|^2 B_0.
\label{eq:B}
\end{equation}  
In the absence of shear ($a=0, x'=x$), these two equations
reduce to the case studied by Manneville and Pomeau \cite{re:manneville83b},
and by Tesauro and Cross \cite{re:tesauro87}.

Finally, we let $\epsilon^{1/2} A_0 =r_A e^{i \phi_A}$, 
$\epsilon^{1/2} B_0 =r_B e^{i \phi_B}$ where to lowest order 
$\phi_A$ and $\phi_B$ are independent of $X$ and $\bar{X}$ \cite{fo:fd6_1},
and re-write the set of amplitude equations in the original 
(unscaled) set of variables. The resulting equations for $r_A$ and $r_B$ 
read 
\begin{equation}
\partial_t r_A = 4 q_0^2 \partial^2_{x'} r_A + (\epsilon - q_0^4 a^4) r_A
             -r_A^3 -2 r_B^2 r_A,
\label{eq:rA}
\end{equation}
and, 
\begin{equation}
\partial_t r_B = - \partial^4_{x'} r_B + 4 a^2 q_0^2 \partial^2_{x'} r_B  +
              \epsilon r_B -r_B^3 -2 r_A^2 r_B. 
\label{eq:rB}
\end{equation}
This is one of the main results of our calculation. At this order, there are
two contributions to the amplitude equation arising from the shear. One in
eq (\ref{eq:rB}) which multiplies the second normal derivative of the 
amplitude $r_{B}$. This contribution is nonzero only in the grain boundary
region, of twice the frequency of the imposed shear, and describes variations
of the amplitude of the parallel lamellae due to the changing orientation 
of the grain boundary with respect to the lamellar planes. 
As we will show below, this term alone originates
an oscillatory contribution to the velocity of the grain boundary of
zero average. The second contribution is the term $q_{0} a^{4} r_{A}$ in
eq (\ref{eq:rA}). This term does not contain a spatial derivative, and
therefore is important in the entire bulk region A where the amplitude
$r_{A}$ does not vanish. It leads to a change in the amplitude of the 
uniform transverse lamellae as they are advected by the flow. 
The corresponding change in the free energy of region A 
does not average to zero over a period and is responsible 
for the net motion of the grain boundary, as shown below.

The amplitude equations (\ref{eq:rA}) and (\ref{eq:rB}) need to be 
supplemented with appropriate boundary conditions. First we have that
$r_A (- \infty,t) =0$ and $r_B (+ \infty,t) =0$.
Furthermore, at large distances from the grain boundary inside domain B, 
$r_B$ reduces to the constant $\sqrt\epsilon$, independent of the 
flow parameters. By contrast, the amplitude $r_A$  inside domain A 
satisfies the equation $\partial_t r_A = (\epsilon - q_0^4 a^4) r_A -r_A^3$
in the limit $x' \rightarrow + \infty$. That equation admits the solution
\begin{equation} 
r_A(+\infty,t)= \{ e^{- f(t)} [ 2 \int_0^t e^{f(t')} dt' + 
      r_A^{-2}(+\infty,0)] \}^{-1/2},
\label{eq:ra_asymp}
\end{equation}
where $f(t)= \left( 2 \epsilon - \frac{3}{4} q_0^4 \gamma^4 \right) t 
 +  q_0^4 \gamma^4 \left( \frac{\sin 2 \omega t}{2 \omega} -  
   \frac{\sin 4 \omega t}{16 \omega}  \right)$. The asymptotic behavior 
of eq (\ref{eq:ra_asymp}) at large times changes qualitatively with the 
sign of the constant
$2 \epsilon - \frac{3}{4} q_0^4 \gamma^4$. When this constant is negative, 
the prefactor $e^{- f(t)}$  
diverges exponentially with time and $r_A(+\infty,t)$ decays to zero. 
If, on the other hand, $2 \epsilon - \frac{3}{4} q_0^4 \gamma^4 > 0$,
Laplace's method can be used to approximate the integral in the expression 
for $r_A$, which reduces to a periodic function 
\begin{equation}
r_A(+\infty,t) = \left\{ \sqrt{\frac{\pi}{g(t)}} e^{h^2(t)/4 g(t)}  
{\mbox{erfc}} \left[ \frac{h(t)}{2 \sqrt{g(t)}} \right]
\right\}^{-1/2},
\label{eq:boundary}
\end{equation}
where $g(t) = \omega q_0^4 \gamma^4 (\sin 2 \omega t - \frac{1}{2}
\sin 4 \omega t)$ and $h(t) = 2 \epsilon + q_0^4 \gamma^4 ( \cos 2 \omega t 
- \frac{1}{4} \cos 4 \omega t -\frac{3}{4})$. 
The condition $2 \epsilon - \frac{3}{4} q_0^4 \gamma^4=0$ which separates 
these two cases can be understood in
terms of a maximum strain amplitude $\gamma^*=(8 \epsilon/ 3 q_0^4)^{1/4}$ 
above which the lamellar phase of domain A will melt. Note that 
$\gamma^{*}$ is independent of the shear frequency $\omega$.

We have numerically solved the coupled, one dimensional equations 
(\ref{eq:rA}) and (\ref{eq:rB}).
The results that will be shown correspond to
$\epsilon = 0.04, q_{0} = 1$ and $\gamma = 0.3$, and a variety of 
shear frequencies. In the calculations, region A was surrounded by two 
identical domains of parallel (B) lamellae so that periodic boundary 
conditions in both directions could be used. The equations were
integrated with a pseudo-spectral algorithm described in detail in
ref. \cite{re:drolet99}.
We used a computational domain of size $L = 4096$ and a grid spacing
$\Delta x = 0.5$.
The time interval $\Delta t$ was chosen a function
of the period of the shear as $\Delta t = 2\pi/(50000 \omega)$.
Stationary solutions obtained  in the absence of shear provided
the initial conditions for $r_A$ and $r_B$.
  
The instantaneous location of the grain boundary $x^{*}$ was defined
as $r_B(x^*)=\sqrt{\epsilon}/2$, and its velocity $v_{gb}$ 
as the rate of change of $x^*$.  Figure \ref{fi:vt} shows $v_{gb}$ 
as a function of time for several values of the angular
frequency $\omega$. Time has been scaled by the period $\tau=2 \pi/\omega$ 
of the applied shear. Positive (resp. negative) values of $v_{gb}$ indicate 
motion toward lamellae in domain A (resp. B). Following an initial transient,  
the velocity oscillates in time at half the period of the shear,
with an amplitude that decreases with the frequency. Note also that
the average velocity is positive; i.e., motion is directed toward domain
A. It is possible to further interpret these results by obtaining an analytic 
approximation for the boundary velocity in the limit of very low frequencies.
If the frequency is sufficiently low, the order parameter $\psi$ (or the 
amplitudes $r_{A}$ and $r_{B}$) will adiabatically follow the motion of the 
boundary. In this limit it is possible to invoke a quasi-static approximation
according to which $r_{A}(x',t) \simeq r_{A}^{s}(x'-x_{gb}^{\prime}(t);a)$
and $r_{B}(x',t) \simeq r_{B}^{s}(x'-x_{gb}^{\prime}(t);a)$, where 
$r_{A}^{s}$ and $r_{B}^{s}$ are stationary solutions of eqs (\ref{eq:rA}) and 
(\ref{eq:rB}) (with the boundary conditions given, including 
eq (\ref{eq:boundary})) that still formally depend on the parameter $a$.
This latter dependence results from the dependence of the stationary
amplitude profiles $r_{A}^{s}$ and $r_{B}^{s}$ on the state of shear of 
the system given by the parameter $a$. Then 
$(\partial_t r_A, \partial_t r_B) \simeq  - v_{gb} (\partial_{\tilde x} 
r_{A}^{s}, \partial_{\tilde x} r_{B}^{s})$ where we have introduced the
notation $\tilde{x}=x'-x_{gb}^{\prime}$.
Following Manneville and Pomeau \cite{re:manneville83b}, we  multiply     
eq (\ref{eq:rA}) by $\partial_{\tilde x} r_{A}^{s}$ and eq (\ref{eq:rB}) by 
$\partial_{\tilde x} r_{B}^{s}$, add the two equations and integrate over 
$\tilde x$ to obtain
\begin{equation}
- v_{gb} \int_{-\infty}^{+\infty} d \tilde x ~ [ (\partial_{\tilde x} 
r_{A}^{s})^2
 +  (\partial_{\tilde x} r_{B}^{s})^2 ] = K(+ \infty) - K(- \infty),
\label{eq:vel1}
\end{equation}
where 
\begin{eqnarray}
\nonumber
K(\tilde x) &=& \frac{1}{2} \left\{ (\epsilon - q_0^4 a^4) r_{A}^{s\;2} + 
       \epsilon  r_{B}^{s\;2}  - \frac{1}{2} (r_{A}^{s\;4} + 
       r_{B}^{s\;4}) -2 r_{A}^{s\;2} r_{B}^{s\;2} 
      +  4 q_0^2 ( \partial_{\tilde x} r_{A}^{s})^2  \right. \\  \nonumber
     & & \left.  + 4 a^{2} q_0^2 
     ( \partial_{\tilde x} r_{B}^{s})^2 -2 \left[ ( \partial_{\tilde x}^3 
     r_{B}^{s}) 
  ( \partial_{\tilde x} r_{B}^{s}) -\frac{1}{2} ( \partial_{\tilde x}^2 
  r_{B}^{s})^2
  \right] \right\}.
\end{eqnarray}
The integral in the left hand side of eq (\ref{eq:vel1}) is an
inverse mobility or friction coefficient, whereas the right hand
side is the effective driving force for grain boundary motion. It is
equal to the static free energy increase of the configuration upon
shearing relative to the planar, unshared boundary and
can be evaluated by using the values $r_{A}^{s} (- \infty) = 0, r_{B}^{s} 
(- \infty) = \sqrt{\epsilon}$ and $r_{B}^{s} (+ \infty) = 0$. 
Furthermore in the quasistatic limit
the function $g(t)$ appearing in eq (\ref{eq:boundary})
is small, and 
$r_{A}^{s} \approx \sqrt{\epsilon - q_0^4 a^4}$. 
As a result, $K(- \infty) = \epsilon^2/4,  K(+ \infty) = (\epsilon -
 q_0^4 a^{4})^2/4$ and    
\begin{equation}
v_{gb} = \frac{q_0^4 \gamma^4 \sin^4(\omega t) [2 \epsilon - 
  q_0^4 \gamma^4 \sin^4(\omega t)]} {4 \int_{-\infty}^{+\infty} d \tilde x ~ 
   [ (\partial_{\tilde x} r_{A}^{s})^2 + (\partial_{\tilde x} r_{B}^{s})^2 ]}.
\label{eq:v2}
\end{equation}
We now compare this result with those obtained by numerical integration of
the governing equations and shown in Fig. \ref{fi:vt} for a range of angular 
frequencies. In order to compute the inverse mobility coefficient
that appears in the denominator of eq (\ref{eq:v2}) we have obtained
$r_A(x',t)$ and $r_B(x',t)$ directly by integration of
eqs (\ref{eq:rA}) and (\ref{eq:rB}). Equation (\ref{eq:v2})
agrees very well 
with the numerical value of $v_{gb}$ at very small shear frequencies, 
as illustrated in Figure \ref{fi:vt} for $\omega =0.001$. The agreement 
progressively deteriorates as the angular frequency increases.

We finally note that although the time dependence of $v_{gb}$ 
changes significantly with $\omega$, its average over a period  
$\langle v_{gb} \rangle = (1/\tau) \int_0^\tau v(t) ~ dt$ only shows
a weak dependence on shear frequency, as shown in figure \ref{fi:vomega}.  
We find, for example, that increasing the shear frequency by two orders of 
magnitude causes a decrease of only 5\% in the average speed of the 
grain boundary. This result follows from the fact that the effective
driving force  $K(- \infty) -  K(+ \infty)$ responsible for the motion 
of the grain boundary is independent of the shear frequency in the
quasistatic limit and increases only marginally at larger frequencies.
As a result, variations in velocity with $\omega$ 
arise solely from the small changes of the inverse mobility coefficient
on frequency.  
Somewhat unexpectedly, the quasistatic approximation does quite well
at predicting the average velocity of the grain boundary for a wide
range of frequencies. We show in Fig. \ref{fi:vomega} 
$\langle v_{gb} \rangle$ obtained by averaging both sides of 
eq (\ref{eq:vel1}) over a period, and the numerical results of Fig. 
\ref{fi:vt} also averaged over a period of the shear.

\section{Discussion}

Following a quench of the diblock copolymer from an initially disordered 
configuration at $T > T_{ODT}$ to a final temperature below 
$T_{ODT}$, the following qualitative picture emerges concerning
the asymptotic, long time selection of a lamellar orientation relative to the
shear. In the absence
of shear ($\gamma = 0$), initial composition fluctuations are
amplified exponentially, with a growth rate that is
isotropic. Lamellar regions emerge and coarsen as a function of
time. Coarsening rates and the role of topological defects in a two
dimensional system have been discussed in ref. \cite{re:boyer01b}. 
To our knowledge, a similar investigation in three dimensions has not
been carried out. If the quench takes place under shear, the
mean field instability threshold depends on orientation, as shown in
ref. \cite{re:drolet99}. The first threshold is to a mixed
parallel-perpendicular mode at $B_{c} = 1/4$, followed by a
bifurcation to a transverse mode at $B_{c} = 1/4 - \gamma^{4}/32 +
{\cal O}(\gamma^{6})$, and to a parallel-transverse mode at
$B_{c} = 1/4 - \gamma^{2}/8 + {\cal O} (\gamma^{4})$. Therefore for
shallow quenches fluctuations along different orientations would be amplified
at different rates leading to predominantly parallel and
perpendicular oriented domains even from an isotropic initial condition. 
Thermal fluctuations, on
the other hand, are known to significantly modify these conclusions
\cite{re:cates89,re:bates90,re:fredrickson94,re:hohenberg95}. In
particular, thermal fluctuations render the mean field supercritical
bifurcation a weakly subcritical bifurcation, with a transition
temperature that increases with $\gamma$. In any event,
the distribution of orientations following a quench in
shear flow is expected not to be isotropic. 

Regardless of whether the copolymer is quenched in shear flow or not,
a macroscopically disordered configuration will result at intermediate
times comprising regions of well saturated monomer composition, but
with a wide distribution of lamellar orientations. The distribution of
observable orientations is reduced by the shear, as those orientations
that are unstable against long wavelength perturbations will quickly
decay when the monomer composition locally reaches nonlinear
saturation. Insofar the melt is Newtonian at the low shear 
frequencies investigated in
refs. \cite{re:drolet99,re:chen02}, one would expect lamellae that are
predominantly perpendicular, and to a lesser degree parallel,
with small projections of the
lamellar wavevector on the transverse direction. There also
exists a small region of stable transverse lamellae. 
Further structure coarsening under shear will involve an initially anisotropic 
distribution of orientations, and hence is expected to be qualitatively 
different from the isotropic case. In metals, for example, coarsening of an 
initially anisotropic distribution of orientations leads to texture 
\cite{re:mullins93}.

The results of this paper further indicate that regions
where parallel and transverse lamellae meet will move, even when both parallel 
and transverse lamellae are linearly stable. The net
motion of the grain boundary is driven by free energy reduction because
parallel lamellae are unaffected by the shear, whereas transverse
lamellae are elastically compressed, a compression that leads to an
increase in energy that is relieved through boundary motion. Since a
similar argument can be made for a boundary separating perpendicular
and transverse lamellae, we would expect that regions of a macroscopic
sample oriented along any combination of parallel and perpendicular
orientations will grow at the expense of any remaining transverse
lamellae.

We are then confronted with an interesting question concerning the
behavior of boundaries separating parallel and perpendicular
lamellae. Both orientations show fluid like response to the shear, in contrast
with the (one dimensional) elastic response of the transverse orientation. 
In fact, the flow does not couple to the monomer composition
for the simple model of Newtonian melt with constant viscosity adopted here, 
and hence there are no shear flow effects on this type of boundary.

If viscous or viscoelastic contrast between the microphases is
allowed, a secondary flow appears which is orientation dependent
\cite{re:fredrickson94,re:chen02}. The velocity field of this
secondary flow is parallel to the lamellar planes (assuming
incompressibility), and largest for a uniform parallel configuration,
while it vanishes for a uniform perpendicular configuration. This
flow is weak in the weak segregation limit, and has negligible
consequences on the stability of a lamellar configuration against long
wavelength perturbations \cite{re:chen02}. However, its possible
effect on boundary motion has not been investigated yet. For long
wavelength modulations of the type described by amplitude equations, it
is possible that the effect of these secondary flows can be subsumed
into an effective constitutive relation for the dissipative part of
the stress tensor as a function of the velocity gradient tensor. This
constitutive relation has to be compatible with the uniaxial symmetry of a 
lamellar phase, in analogy with other uniaxial systems such as
nematic or smectic liquid crystals. Additional viscosity coefficients would 
enter the constitutive law, as well as an explicit dependence on ${\mathbf q}$,
the slowly varying normal to the lamellar planes. In this case, the
effective viscosity of a region of parallel lamellae is different from 
that of perpendicular lamellae thus leading to secondary flows that are
asymmetric with respect to the boundary, and possibly to boundary motion. 
This possibility is currently under investigation.

\section*{Acknowledgments} This research has been supported by the
National Science Foundation under contract DMR-0100903.

\bibliographystyle{achemso}
\bibliography{$HOME/mss/references}

\providecommand{\refin}[1]{\\ \textbf{Referenced in:} #1}
\begin{thebibliography}{10}

\bibitem{re:boyer01}
Boyer,~D.;\ \ {Vi\~nals},~J. \textit{Phys. Rev. E} \textbf{2001,} \textsl{63,}
  061704.

\bibitem{re:fasolka97}
Fasolka,~M.;\ \ Harris,~D.;\ \ Mayes,~A.;\ \ Yoon,~M.;\ \ Mochrie,~S.~G.
  \textit{Phys. Rev. Lett.} \textbf{1997,} \textsl{79,} 3018.

\bibitem{re:rockford99}
Rockford,~L.;\ \ Liu,~Y.;\ \ Mansky,~P.;\ \ Russell,~T.;\ \ Yoon,~M.;\ \
  Mochrie,~S. \textit{Phys. Rev. Lett.} \textbf{1999,} \textsl{82,} 2602.

\bibitem{re:fasolka01}
Fasolka,~M.;\ \ Hayes,~A. \textit{Annu. Rev. Mater. Sci.} \textbf{2001,}
  \textsl{31,} 323.

\bibitem{re:segalman01}
Segalman,~R.;\ \ Yokoyama,~H.;\ \ Kramer,~E. \textit{Adv. Mater.}
  \textbf{2001,} \textsl{13,} 1152.

\bibitem{re:amundson93}
Amundson,~K.;\ \ Helfand,~E.;\ \ Quan,~X.;\ \ Smith,~D. \textit{Macromolecules}
  \textbf{1993,} \textsl{26,} 2698.

\bibitem{re:amundson94}
Amundson,~K.;\ \ Helfand,~E.;\ \ Quan,~X.;\ \ Hudson,~D.;\ \ Smith,~D.
  \textit{Macromolecules} \textbf{1994,} \textsl{27,} 6559.

\bibitem{re:tsori03}
Tsori,~Y.;\ \ Tournilhac,~F.;\ \ Leibler,~L. arXiv:cond-mat/0301116.

\bibitem{re:hadziiaannou79}
Hadziiaannou,~G.;\ \ Mathis,~A.;\ \ Skoulios,~A. \textit{Colloid Polym. Sci.}
  \textbf{1979,} \textsl{257,} 136.

\bibitem{re:koppi92}
Koppi,~K.;\ \ Tirrell,~M.;\ \ Bates,~F.;\ \ Almdal,~K.;\ \ Colby,~R. \textit{J.
  Phys. II (France)} \textbf{1992,} \textsl{2,} 1941.

\bibitem{re:koppi93}
Koppi,~K.;\ \ Tirrel,~M.;\ \ Bates,~F. \textit{Phys. Rev. Lett.} \textbf{1993,}
  \textsl{70,} 1449.

\bibitem{re:patel95}
Patel,~S.;\ \ Larson,~R.;\ \ Winey,~K.;\ \ Watanabe,~H. \textit{Macromolecules}
  \textbf{1995,} \textsl{28,} 4313.

\bibitem{re:gupta95}
Gupta,~V.;\ \ Krishnamoorti,~R.;\ \ Kornfield,~J.;\ \ Smith,~S.
  \textit{Macromolecules} \textbf{1995,} \textsl{28,} 4464.

\bibitem{re:leist99}
Leist,~H.;\ \ Maring,~D.;\ \ Thurn-Albrecht,~T.;\ \ Wiesner,~U. \textit{J.
  Chem. Phys.} \textbf{1999,} \textsl{110,} 8225.

\bibitem{re:hamley98}
Hamley,~I. \textit{The Physics of Block Copolymers;} Oxford University Press:
  New York, 1998.

\bibitem{re:larson99}
Larson,~R.~G. \textit{The Structure and Rheology of Complex Fluids;} Oxford
  University Press: New York, 1999.

\bibitem{re:drolet99}
Drolet,~F.;\ \ Chen,~P.;\ \ {Vi\~nals},~J. \textit{Macromolecules}
  \textbf{1999,} \textsl{32,} 8603.

\bibitem{re:chen02}
Chen,~P.;\ \ {Vi\~nals},~J. \textit{Macromolecules} \textbf{2002,} \textsl{35,}
  4183.

\bibitem{re:zvelindovsky98}
Zvelindovsky,~A.;\ \ Sevink,~G.;\ \ {van Vlimmeren},~B.;\ \ Maurits,~N.;\ \
  Fraaije,~J. \textit{Phys. Rev. E} \textbf{1998,} \textsl{57,} R4879.

\bibitem{re:ren01}
Ren,~S.;\ \ Hamley,~I.;\ \ Teixeira,~P.;\ \ Olmsted,~P. \textit{Phys. Rev. E}
  \textbf{2001,} \textsl{63,} 041503.

\bibitem{re:leibler80}
Leibler,~L. \textit{Macromolecules} \textbf{1980,} \textsl{13,} 1602.

\bibitem{re:ohta86}
Ohta,~T.;\ \ Kawasaki,~K. \textit{Macromolecules} \textbf{1986,} \textsl{19,}
  2621.

\bibitem{re:oono88b}
Oono,~Y.;\ \ Bahiana,~M. \textit{Phys. Rev. Lett.} \textbf{1988,} \textsl{61,}
  1109.

\bibitem{re:swift77}
Swift,~J.;\ \ Hohenberg,~P. \textit{Phys. Rev. A} \textbf{1977,} \textsl{15,}
  319.

\bibitem{re:cates89}
Cates,~M.;\ \ Milner,~S. \textit{Phys. Rev. Lett.} \textbf{1989,} \textsl{62,}
  1856.

\bibitem{re:fredrickson94}
Fredrickson,~G. \textit{J. Rheol.} \textbf{1994,} \textsl{38,} 1045.

\bibitem{re:shiwa97}
Shiwa,~Y. \textit{Physics Letters A} \textbf{1997,} \textsl{228,} 279.

\bibitem{re:boyer02}
Boyer,~D.;\ \ {Vi\~nals},~J. \textit{Phys. Rev. E} \textbf{2002,} \textsl{65,}
  046119.

\bibitem{re:tesauro87}
Tesauro,~G.;\ \ Cross,~M. \textit{Phil. Mag. A} \textbf{1987,} \textsl{56,}
  703.

\bibitem{re:manneville83b}
Manneville,~P.;\ \ Pomeau,~Y. \textit{Phil. Mag. A} \textbf{1983,} \textsl{48,}
  607.

\bibitem{fo:fd6_1}
In the amplitude equation formalism the phase of the solution is also
  independent of the fast variable $x'$; {\em i.e.} the location of the grain
  boundary is independent of the phase of the periodic, base solution. This
  independence is broken when non adiabatic effects are incorporated, as shown
  in ref. \protect\cite{re:boyer02}. Sufficiently close to threshold, non
  addiabatic corrections to the amplitude equation of a lamellar phase are
  small.

\bibitem{re:boyer01b}
Boyer,~D.;\ \ {Vi\~nals},~J. \textit{Phys. Rev. E} \textbf{2001,} \textsl{64,}
  050101(R).

\bibitem{re:bates90}
Bates,~F.;\ \ Fredrickson,~G. \textit{Annu. Rev. Phys. Chem.} \textbf{1990,}
  \textsl{41,} 525.

\bibitem{re:hohenberg95}
Hohenberg,~P.;\ \ Swift,~J. \textit{Phys. Rev. E} \textbf{1995,} \textsl{52,}
  1828.

\bibitem{re:mullins93}
Mullins,~W.;\ \ {Vi\~nals},~J. \textit{Acta metall. mater.} \textbf{1993,}
  \textsl{41,} 1359.

\end{thebibliography}

\newpage
\begin{center}
FIGURE CAPTIONS
\end{center}

\begin{figure}
\centerline{\epsfig{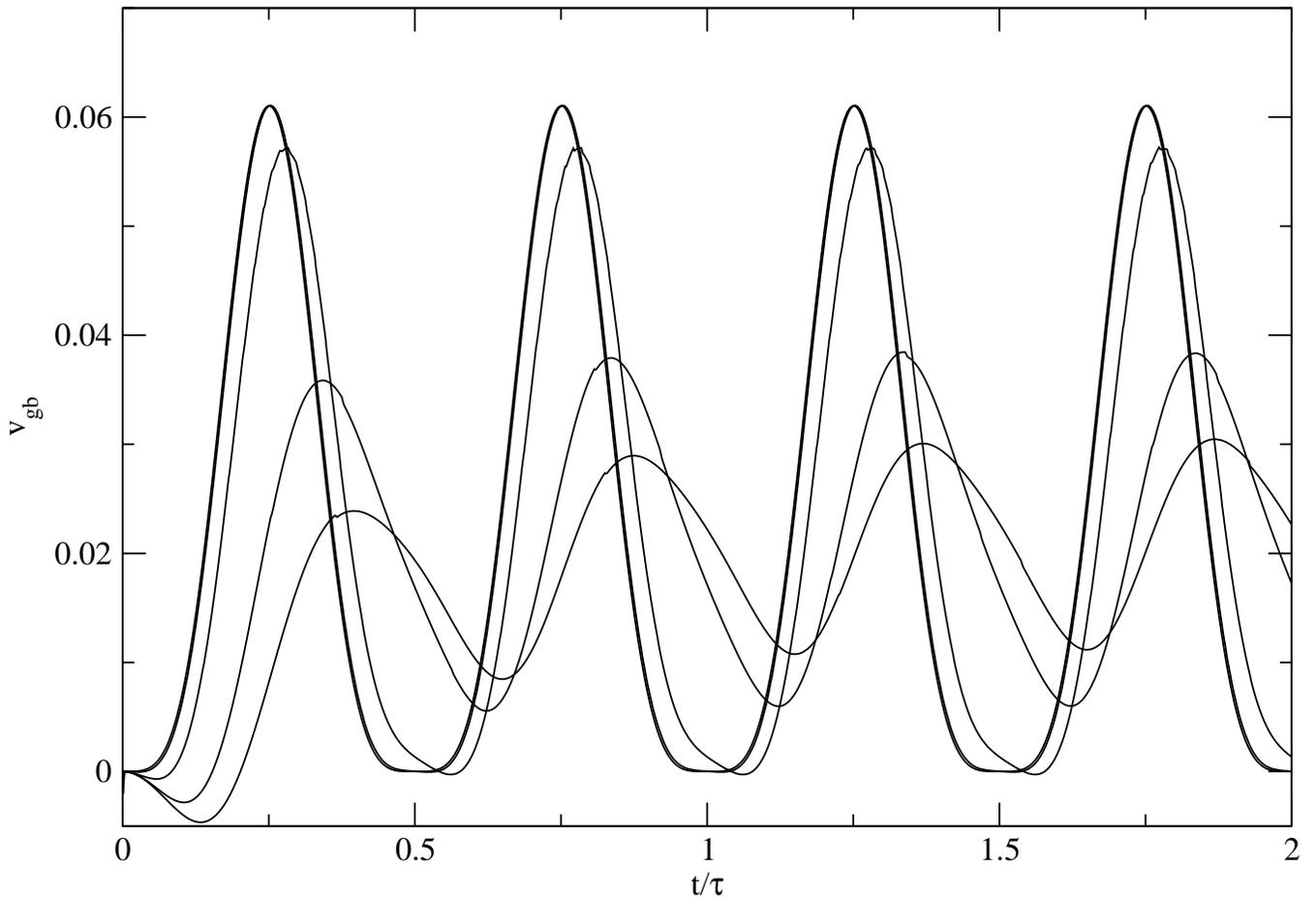}}
\caption{Grain boundary velocity as a function of time obtained by 
numerical solution of eqs (\ref{eq:A1}) and (\ref{eq:B1}). Four different
angular frequencies are shown: (in order of decreasing amplitude)
$\omega = 0.001, 0.01, 0.05,$ and 0.1. Also shown is the quasistatic
approximation of eq (\ref{eq:v2}) calculated at the lowest angular
frequency $\omega = 0.001$. The curve is indistinguishable in the graph
from the corresponding numerical solution.}
\label{fi:vt}
\end{figure}

\begin{figure}
\centerline{\epsfig{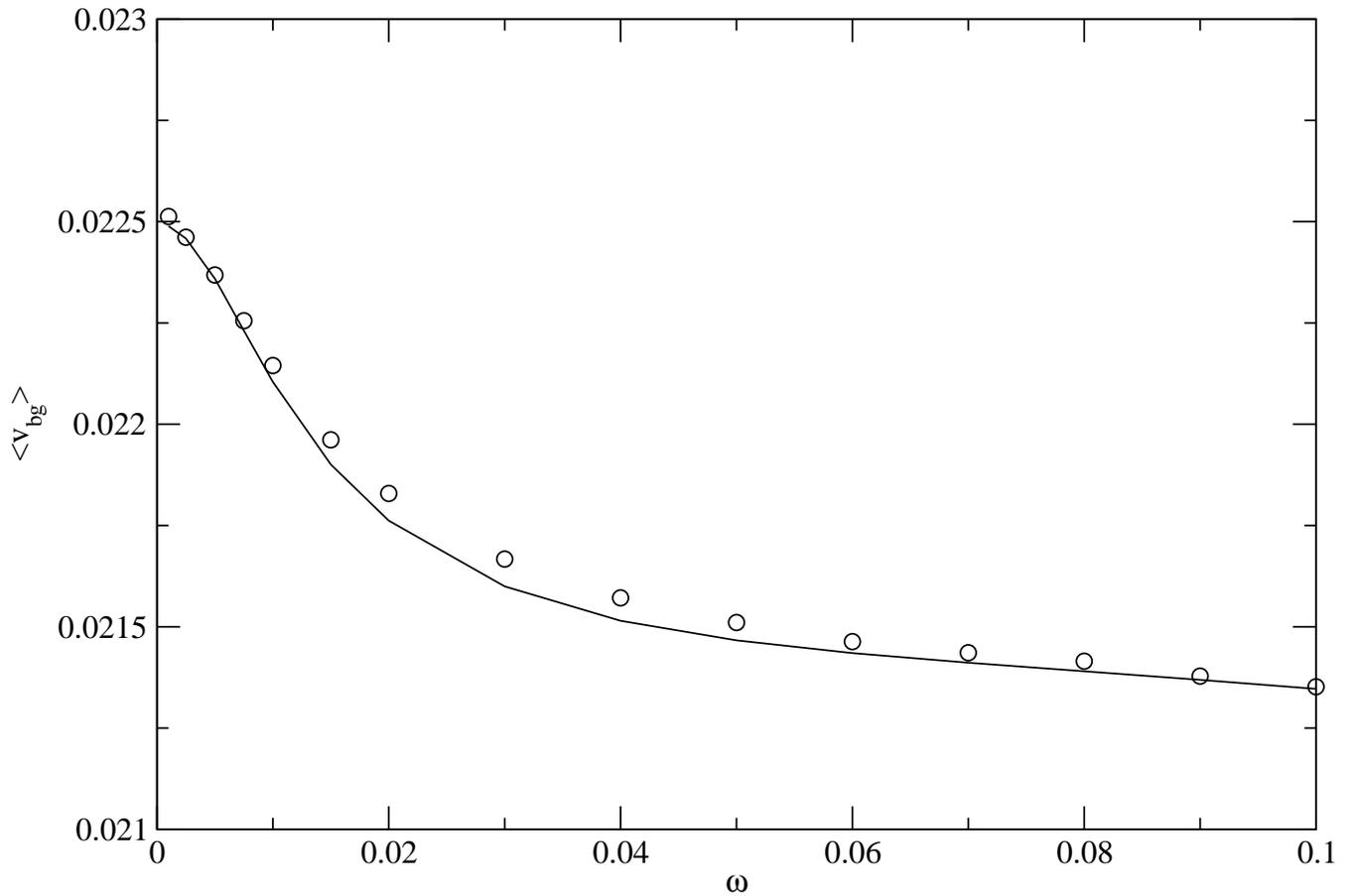}}
%
%
\caption{Temporal average of the grain boundary velocity as a function of the
angular frequency of the shear. The symbols correspond to the time average
of the numerically obtained velocities shown in Fig. \ref{fi:vt}, and the
solid line is the time average of the quasistatic velocity given in 
eq (\ref{eq:v2}).}
\label{fi:vomega}
\end{figure}

\begin{itemize}
\item[Figure 1.] Schematic representation of the geometry considered including
the shear direction, and the three different lamellar orientations discussed
in the text.

\item[Figure 2.] Schematic representation of a planar grain boundary that 
separates regions of parallel and transverse lamellae being uniformly sheared.

\item[Figure 3.] Grain boundary velocity as a function of time obtained by 
numerical solution of eqs (\ref{eq:A1}) and (\ref{eq:B1}). Four different
angular frequencies are shown: (in order of decreasing amplitude)
$\omega = 0.001, 0.01, 0.05,$ and 0.1. Also shown is the quasistatic
approximation of eq (\ref{eq:v2}) calculated at the lowest angular
frequency $\omega = 0.001$. The curve is indistinguishable in the graph
from the corresponding numerical solution.

\item[Figure 4.] Temporal average of the grain boundary velocity as a function 
of the angular frequency of the shear. The symbols correspond to the time 
average of the numerically obtained velocities shown in Fig. \ref{fi:vt}, 
and the solid line is the time average of the quasistatic velocity given in 
eq (\ref{eq:v2}).

\end{itemize}

\newpage

~

\newpage

\newpage

\newpage


\end{document}